\documentstyle[12pt]{article}
\newlength{\extraspace}
\newlength{\extraspaces}
\def\bsklength{2mm} 

\begin{document}
\addtolength{\baselineskip}{\bsklength}
\thispagestyle{empty}

\begin{flushright}
{\sc CTP-TAMU}-84/92\\
hep-ph/9212286\\
December 1992
\end{flushright}
\vspace{.6cm}

\begin{center}
{\LARGE{{Electroweak Z-string in \\[3mm]
Two-Higgs-Doublet Standard Model}
}}\\[1.5cm]
 {\large HoSeong La}%
\footnote{e-mail address: hsla@phys.tamu.edu, hsla@tamphys.bitnet}\\[3mm]
{\it Center for Theoretical Physics\\[2mm]
Texas A\&M University\\[2mm]
College Station, TX 77843-4242} \\[2cm]

{\sc Abstract}\\[1cm]
{\parbox{15cm}{
\addtolength{\baselineskip}{\bsklength}
We derive electroweak Z-string solutions in the Glashow-Weinberg-Salam model
with two Higgs doublets. The existence of such solutions in particular requires
a specific relation between  the
ratio of the two Higgs vacuum expectation values, {\it i.e.}
 $\tan\beta$, and the couplings in the Higgs potential.
}
}
\end{center}
\bigskip\bigskip
{\noindent PACS numbers: 12.15.-y, 11.15.Ex, 11.30.Er, 11.17.+y }

\noindent
\vfill

\def\ie{{\it i.e.\ }}
\def\eg{{\it e.g.\ }}
\def\VEV{VEV}
\renewcommand{\hat}{\widehat}
\renewcommand{\det}{{\rm det}}
\def\la{\lambda}
\def\hf{{1\over 2}}
\def\dag{\dagger}
\newcommand{\e}{{\rm e}}
\newcommand{\eps}{\epsilon}
\newcommand{\lbr}{\left(}
\newcommand{\rbr}{\right)}
\newfont{\cmss}{cmss10 scaled 1200} \newfont{\cmsss}{cmss10 scaled 833}
\def\IZ{\relax\ifmmode\mathchoice
{\hbox{\cmss Z\kern-.4em Z}}{\hbox{\cmss Z\kern-.4em Z}}
{\lower.9pt\hbox{\cmsss Z\kern-.4em Z}}
{\lower1.2pt\hbox{\cmsss Z\kern-.4em Z}}\else{\cmss Z\kern-.4em Z}\fi}
\def\IR{\relax\ifmmode\mathchoice
{\hbox{\cmss I\kern-.5em I}}{\hbox{\cmss R\kern-.5em R}}
{\lower.9pt\hbox{\cmsss I\kern-.5em I}}
{\lower1.2pt\hbox{\cmsss R\kern-.5em R}}\else{\cmss R\kern-.5em R}\fi}

\newcommand{\beq}{\begin{equation}
\addtolength{\abovedisplayskip}{\extraspaces}
\addtolength{\belowdisplayskip}{\extraspaces}
\addtolength{\abovedisplayshortskip}{\extraspace}
\addtolength{\belowdisplayshortskip}{\extraspace}}
\newcommand{\eeq}{\end{equation}}

\newcommand{\beqa}{\begin{eqnarray}
\addtolength{\abovedisplayskip}{\extraspaces}
\addtolength{\belowdisplayskip}{\extraspaces}
\addtolength{\abovedisplayshortskip}{\extraspace}
\addtolength{\belowdisplayshortskip}{\extraspace}}
\newcommand{\eeqa}{\end{eqnarray}}

\newcommand{\Tr}{{\rm Tr}}
\newcommand{\tr}{{\rm tr}}
\newcommand{\nonu}{\nonumber \\[.5mm]}
\def\half{\textstyle{1\over 2}}
\newcommand{\pa}{\partial}
\newcommand{\pab}{\overline{\partial}}
\newcommand{\zb}{{\overline{z}}}
\newcommand{\thetabar}{{\overline{\theta}}}
\newcommand{\from}{\leftarrow}
\newcommand{\rra}{\ \longrightarrow \ }
\newcommand{\lla}{\ \longleftarrow\ }
\newcommand{\mod}[1]{\quad (\mbox{mod}\;#1)}
\newcommand{\llrra}{{\mathop{\longleftrightarrow}}}
\newcommand{\lra}{{\mathop{\leftrightarrow}}}
\newcommand{\CH}{{\cal H}}
\newcommand{\CZ}{{\cal Z}}
\newcommand{\CP}{{\cal P}}
\newcommand{\CO}{{\cal O}}
\newcommand{\CS}{{\cal S}}
\newcommand{\CL}{{\cal L}}
\def\lbr{\left(}
\def\rbr{\right)}
\newcommand{\psibar}{{\overline{\psi}}}
\newcommand{\wtilde}{\widetilde}
\def\gt{\widetilde{g}}
\newcommand{\Ph}{{\mbox{\raisebox{.1ex}{$\phi$}}}}
\newcommand{\rg}{{\mbox{\raisebox{.1ex}{$g$}}}}
\newcommand{\lfr}[2]{{\textstyle {#1\over #2}}}

\newpage

\newcounter{xxx}
\setlength{\parskip}{2mm}
\addtolength{\baselineskip}{\bsklength}



One of the most mysterious parts of the electroweak theory 
lies in the Higgs
sector. Higgs was introduced to achieve the electroweak symmetry breaking
without spoiling the consistency of the theory. 
However, this so-much-wanted scalar particle has been
escaping from all current searches and is still at large.
The spontaneous symmetry breaking induced by the Higgs system can
often generate vacuum defects\cite{Vilrev}\
and it turns out that the electroweak theory may
not be an exception\cite{Vach}.

In this letter we shall investigate the structure of a string-like defect
(so-called ``Z-string")
in the two-Higgs-doublet standard model\cite{rtwoH,revHig}.
This is also strongly
motivated by the recent growing anticipation that the minimal supersymmetric
Grand Unified Theories (GUT's) may lead to
a phenomenologically plausible unified
theory of strong and electroweak interactions.
These supersymmetric GUTs in general require at least two Higgs
multiplets for the electroweak symmetry breaking\cite{revSGUT}.
Thus the true Higgs system to lead to the electroweak symmetry breaking
may be a multi-Higgs one.

In the two-Higgs-doublet models
each Higgs gets its own vacuum expectation value (\VEV), say $v_1, v_2$,
to spontaneously break the $SU(2)\times U(1)_Y$ symmetry
down to the $U(1)_{em}$. These \VEV s are phenomenologically
important  but unfortunately they are not determined
theoretically except in some no-scale models\cite{rNano}.
The geometric sum $v^2/2=v_1^2+v_2^2$ can be determined in terms of the
mass of the gauge boson, where $v$ denotes
the electroweak symmetry breaking scale.
This however leaves the ratio of the two \VEV s,
$\tan\beta\equiv v_2/v_1$,  still
undetermined. Thus it is very important to understand
the {\it rationale} behind the symmetry breaking with two \VEV s
and to look for any mechanism
to determine the ratio rather theoretically, if possible.

With such motivation in mind we shall pay particular attention to the role of
$\tan\beta$ in the structure of the Z-string. A toy model of
two Higgs scalars coupled to the $U(1)$ abelian gauge field has been
investigated by the author before and indeed vortex solution in this model
requires a specific relation between
$\tan\beta$ and the couplings in the Higgs potential\cite{mhiggs}.
In this letter we report that such a structure is indeed quite generic
even in a realistic model like the standard model, although there are
some subtle differences involved. We shall present the backbone of the
structure here, but more detail can be found in ref.\cite{dhiggs}.
We also expect that such a structure will persist in the supersymmetric cases.

We shall use the CP invariant two-doublet Higgs potential that
induces $SU(2)\times U(1)_Y\to U(1)_{em}$ symmetry breaking\cite{GerH,revHig}:
\beq\label{edhigpot}
\begin{array}{cl}
V(\phi_1,\phi_2)=&\lfr{1}{2}\la_1\left(|\phi_1|^2-v_1^2\right)^2+
\lfr{1}{2}\la_2\left(|\phi_2|^2-v_2^2\right)^2+
\lfr{1}{2}\la_3\left(|\phi_1|^2+|\phi_2|^2-v_1^2-v_2^2\right)^2 \\ [5mm]
&+\la_4\left(|\phi_1|^2 |\phi_2|^2-|\phi_1^\dag\phi_2|^2\right)
+\la_5\left|\phi_1^\dag\phi_2 -v_1 v_2\right|^2,
\end{array}
\eeq
where $\phi_1,\ \phi_2$ are $SU(2)$ doublets.
In this letter we shall stick to the general case that $\la_i\neq 0$
for $i=1,2,3$ and also assume that all $\la_j,\ j=1,\ldots, 5$ are nonnegative.
This potential shows $\phi_1\leftrightarrow\phi_2$ discrete symmetry, which is
necessary to suppress the flavor changing neutral current.
Then we shall find that this system reveals a rather interesting result,
which cannot be obtained otherwise.
The key observation is that the spontaneous symmetry breaking of
Eq.(\ref{edhigpot})
leads to a vortex solution, whose existence will introduce an extra condition
on the Higgs \VEV s.

Consider the bosonic sector of the standard model
described by the Lagrangian density
\beq\label{edlagr}
{\CL=-\half\tr G_{\mu\nu}G^{\mu\nu}-{\textstyle{1\over 4}}F^{\mu\nu}F_{\mu\nu}
+|D_\mu\phi_1|^2+|D_\mu\phi_2|^2
	-V(\phi_1, \phi_2),}
\eeq
where $F_{\mu\nu}=\pa_\mu B_\nu-\pa_\nu B_\mu$,
$G_{\mu\nu}^a=\pa_\mu W_\nu^a-\pa_\nu W_\mu^a + g\eps^{abc} W_\mu^b W_\nu^c$,
and $D_\mu=\pa_\mu-ig'{Y\over 2}B_\mu-ig{\tau^a\over 2} W_\mu^a$.
Both Higgs' have hypercharge $Y=1$.

Then the equations of motion for the scalar fields are
\beqa
\label{edomia}
&&0=D^\mu D_\mu\phi_1 +\la_1\left(|\phi_1|^2-v_1^2\right)\phi_1
	+\la_3\left(|\phi_1|^2+|\phi_2|^2-v_1^2-v_2^2\right)\phi_1\nonu
&&\quad +\la_4\left(|\phi_2|^2\phi_1-(\phi_2^\dag\phi_1)\phi_2\right)
	+\la_5(\phi_2^\dag\phi_1-v_1 v_2)\phi_2,\\
\label{edomib}
&&0=D^\mu D_\mu\phi_2 +\la_2\left(|\phi_2|^2-v_2^2\right)\phi_2
	+\la_3\left(|\phi_1|^2+|\phi_2|^2-v_1^2-v_2^2\right)\phi_2 \nonu
&&\quad +\la_4\left(|\phi_1|^2\phi_2-(\phi_1^\dag\phi_2)\phi_1\right)
	+\la_5(\phi_1^\dag\phi_2-v_1 v_2)\phi_1,
\eeqa
and for the gauge fields we have
\beqa
\label{edomic}
&&-\pa^\mu F_{\mu\nu}=j_\nu\equiv j_{1\nu} + j_{2\nu}, \\
&&\quad j_{i\nu} \equiv\lfr{1}{2}ig' \left[\phi_i^\dag\pa_\nu\phi_i
-(\pa_\nu \phi_i)^\dag\phi_i\right]
+\lfr{1}{2}g'^2 B_\nu |\phi_i|^2+\lfr{1}{2}g'g W_\nu^a\phi_i^\dag\tau^a\phi_i,
\ \ i=1,2, \nonu
\label{edomid}
&&-\pa^\mu G^a_{\mu\nu} -g\eps^{abc}W^{b\mu}G^c_{\mu\nu}
=J^a_\nu\equiv J^a_{1\nu} + J^a_{2\nu}, \\
&&\quad J^a_{i\nu} \equiv\lfr{1}{2} ig \left[\phi_i^\dag\tau^a\pa_\nu\phi_i
-(\pa_\nu \phi_i)^\dag\tau^a\phi_i\right]
+\lfr{1}{2}gg' B_\nu \phi_i^\dag\tau^a\phi_i
+\lfr{1}{2}g^2W_\nu^b\phi_i^\dag\tau^a\tau^b\phi_i,
\ \ i=1,2, \nonumber
\eeqa
For time-independent solutions we choose $B_0=0=W_0^a$ gauge and impose the
cylindrical symmetry around the string, then the system
effectively reduces to a two-dimensional one. In this case the string solutions
in the $(1+3)$-dimensional spacetime correspond to
the vortex solutions in $\IR^2$.
When Higgs gets \VEV, the false vacuum region forms vacuum defects.
As usual,  we redefine the neutral gauge fields as
\beq\label{azfld}
A_\mu=\cos\theta_W B_\mu+\sin\theta_W W_\mu^3,\quad
Z_\mu=\sin\theta_W B_\mu-\cos\theta_W W_\mu^3,
\eeq
where $\theta_W$ is the Weinberg angle defined by $\tan\theta_W=g'/g$.
We shall also use $\gt\equiv \half\sqrt{g^2+g'^2}$ for convenience.

For vortex solutions
it is convenient to represent them in the polar coordinates
$(r, \theta)$\cite{NieOl}\ such as
\beq\label{eans}{\phi_1=\pmatrix{0\cr\e^{im\theta} f_1(r)\cr},
\quad \  \phi_2=\pmatrix{0\cr \e^{in\theta} f_2(r)\cr},\quad \
	{\vec Z}={\hat e_\theta} {1\over r}Z(r),}
\eeq
where $m,n$ are integers identifying each ``winding'' sector ( we shall
come back to this point later again.).
Here we are mainly interested in the case of $W_\mu^1=0=W_\mu^2$,
but we expect there are other solutions
similar to the case of ref.\cite{Vach}.
Then Eqs.(\ref{edomic},\ref{edomid}) become
\beqa
\label{edmiia}
-{1\over r}\pa_r(r\pa_r B_\theta)+{1\over r^2}B_\theta
-{g'\over r}\left[\left(m-\half(g'B_\theta-g W_\theta^3)\rbr f_1^2
+ (n-\half(g'B_\theta-g W_\theta^3) )f_2^2\right] &=0,\\
\label{edmiib}
-{1\over r}\pa_r(r\pa_r W^3_\theta)+{1\over r^2}W^3_\theta
+{g\over r}\left[\left(m-\half(g'B_\theta-g W_\theta^3)\rbr f_1^2
+ (n-\half(g'B_\theta-g W_\theta^3) )f_2^2\right] &=0.
\eeqa
As we can easily see, $A_\mu$ satisfies a trivial equation so that we can
set $A_\mu=0$. Thus from the rest of the equations of motion we obtain
\beqa
\label{edomiia}
-{1\over r}\pa_r(r\pa_r f_1)\! +{1\over r^2}f_1(m\!-\!\gt Z)^2
\! + \!(\la_1 \!+\!\la_3) (f_1^2\!-\! v_1^2)f_1
\!+\!\la_3 (f_2^2\!-\! v_2^2)f_1 \!+
\!\la_5\!\left(f_1f_2\!- \! v_1v_2\e^{i(n-m)\theta}\right)\! f_2
\!\!\! &\!\!=\! 0,\;\,\,\\
\label{edomiib}
-{1\over r}\pa_r(r\pa_r f_2)\! +{1\over r^2}f_2(n\!-\!\gt Z)^2
\! +\! (\la_2 \!+\!\la_3) (f_2^2\!-\! v_2^2)f_2
\!+\!\la_3 (f_1^2\!-\! v_1^2)f_2 \!+
\!\la_5\!\left(f_1f_2\!-\! v_1v_2\e^{i(m-n)\theta}\right)\! f_1
\!\!\! &\!\!=\! 0,\;\,\, \\
\label{edomiic}
-\pa_r^2 Z+{1\over r}Z
-2\gt\left[(m-\gt Z) f_1^2 + (n-\gt Z)f_2^2\right]\!\!\! &\!\! =\!0. \;\,\,\
\eeqa
Note that $\la_4$ coupling does not take part in this structure classically.

To become  desired finite-energy defects located at $r=0$
the solutions we are looking for should satisfy the
following boundary conditions:
\beq\label{ebc}\begin{array}{cl}
f_1(0)=0,\ \ f_2(0)=0,\ \ &Z(0)=0,\\ [5mm]
f_1\to v_1,\ f_2\to v_2, \ &Z\to {\rm const.}\ \ {\rm as}\ \
r\to\infty.
\end{array}
\eeq
The constant for the asymptotic value of $Z$ will be determined properly later.

In general for arbitrary coupling constants
it will be a formidable task to solve these equations exactly due
to the complexity of the Higgs potential,
but we can always look for asymptotic solutions.
Fortunately, for our purpose it turns out to be
good enough to find approximate solutions for large $r$.

Imposing the boundary conditions at large $r$,
Eqs.(\ref{edomiia},\ref{edomiib})
become consistent only if $m=n$ and that it fixes the asymptotic value
$ Z\to {n/ \gt}$ as $r\to\infty$.
This implies that there is no vortex solution of different ``winding''
numbers for different Higgs fields.
With this condition of winding numbers we can solve
Eq.(\ref{edomiic}) for large $r$ to obtain\cite{NieOl}
\beq\label{dsolA}
Z\to {n\over \gt}-n\sqrt{{\pi v\over 2\gt}}{\sqrt r} \e^{-r/\la}+\cdots,
\eeq
where $\la=1/\gt v$ is the characteristic length of the gauge field.
Note that the characteristic length defines the region over which the field
becomes significantly different from the value at the location of the defect.

The asymptotic solutions for $\phi_1$ and $\phi_2$  can be found as follows:
For simplicity we consider $n=1$ case, but the result does not
really depend on $n$.
Asymptotically we look for solutions of the form
\beq\label{eI}
f_1 -v_1\sim c_1\e^{-r/\xi_1},\ \ f_2-v_2 \sim c_2\e^{-r/\xi_2},
\eeq
where $\la_1$ and $\la_2$ are the characteristic lengths of
$\phi_1$ and $\phi_2$ respectively and
the constant coefficients $c_1$ and $c_2$  are in principle calculable.
Note that  we can normalize any non-dimensionful constants in $c_i$ to be
the same. Furthermore, for our purpose only the ratio is relevant.
Therefore these constants
can be taken as $c_1=-v_1$ and $c_2=-v_2$ in a good approximation.
Then in the leading order we obtain
\beqa
\label{eIIa}
v_1\e^{-r/\xi_1}
\left[-{1\over \xi_1^2}+2(\la_1+\la_3)v_1^2+\la_5 v_2^2
\right] +(2\la_3+\la_5) v_1 v_2^2
\e^{-r/\xi_2}+\cdots &=0 \\
\label{eIIb}
v_2\e^{-r/\xi_2}
\left[-{1\over \xi_2^2}+2(\la_2+\la_3)v_2^2+\la_5 v_1^2
\right] +(2\la_3+\la_5) v_1^2 v_2
\e^{-r/\xi_1}+\cdots &=0,
\eeqa
where the ellipses include terms which vanish more rapidly as $r\to\infty$.

Recall that $\la_3>0$ and $\la_5\geq 0$ so that $2\la_3+\la_5\neq 0$.
Thus to have any vortex solution we are forced to identify the two
characteristic lengths of the scalar fields such that
$$\xi\equiv\xi_1=\xi_2.$$
Therefore we get the desired result by demanding the vanishing coefficients
of $\e^{-r/\xi}$ in Eqs.(\ref{eIIa},\ref{eIIb})\ as
\beq\label{edresult}
\tan\beta\equiv {v_2\over v_1}=\sqrt{{\la_1-\la_5\over
\la_2-\la_5}},\ \ \  \la_3\neq 0\ \ {\rm or} \ \la_5\neq 0.
\eeq
Thus we have determined the ratio of the two Higgs \VEV s in terms of the
couplings in the Higgs potential. This tells us that although different Higgs
field gets different \VEV s, their characteristic lengths should be the same to
form a single defect. Both Higgs should reach the true vacuum at the same
distance. To do that the two \VEV s should satisfy a proper relation, which is
Eq.(\ref{edresult}).

Furthermore, together with $v$, we can completely determine the \VEV s as
\beq\label{eIII}
v_1 ={v\over\sqrt{2}}
\,\cos\beta={v\over\sqrt{2}}{\sqrt{{\la_2-\la_5\over \la_1+\la_2-2\la_5}}},\ \
v_2 ={v\over\sqrt{2}}
\,\sin\beta={v\over\sqrt{2}}{\sqrt{{\la_1-\la_5\over \la_1+\la_2-2\la_5}}}.
\eeq
The characteristic lengths $\xi_1, \xi_2$, now satisfy
\beq\label{echar}
\xi\equiv\xi_1=\xi_2={1\over v}
\sqrt{{\la_1+\la_2-2\la_5\over \la_1\la_2+\la_2\la_3+\la_3\la_1
-\la_5(2\la_3+\la_5)}}.
\eeq
Note that, although $\tan\beta$ does not depend on $\la_3$, it is crucial to
have nonvanishing $\la_3$ or $\la_5$ coupling to obtain such a result.
The gauge boson mass is $M_Z=1/\la=\gt v$
after spontaneous symmetry breaking.

In this two-Higgs-doublet model there are five physical Higgs bosons:
$H^{\pm},\ A^0,\ H^0,\ h^0$. $A^0$ is a CP-odd neutral scalar, while
$H^0,\ h^0$ are CP-even scalars. $h^0$ denotes the lightest Higgs.
Using Eq.(\ref{eIII}), we can compute the masses of all these physical Higgs
bosons in terms of the couplings in the Higgs potential and $v$, where
$v=246$GeV. $\la_5$ is related to $M_{A^0}$ and $M_{H_0,h_0}$ can be determined
in terms of $\la_1,\la_2,\la_3,\la_5$ and $v$. Thus we only have five free
parameters.

The appearance of integers in the solutions,
which we still call ``winding'' number,
is rather intriguing because there is no
explicit $U(1)$ symmetry to be broken
which should determine the necessary topological sector.
If our vortex solutions are nontopological as in ref.\cite{tdlee},
there should not be such a parameter.
This however
can be explained as follows: If we regard $W_\mu^1=0=W_\mu^2$ as gauge fixing
conditions, then effectively we can view
the symmetry of the system as $U(1)\times U(1)_Y$. When we twist this symmetry
to obtain $U(1)_{em}$, the remaining twisted $U(1)_{\gt}$
is spontaneously broken to lead to the winding sector.
This perhaps would be also explained
similarly from the point of view of
ref.\cite{preskill}, which analyzed the topological origin of
the semilocal defects\cite{VaAc}.

So far we have not mentioned anything about the stability of this electroweak
Z-string solution obtained in this model.
Even in the very special case in which the gauge
coupling is related to some of the Higgs couplings,
it is most likely that this solution would not saturate the Bogomol'nyi
bound. Thus it may not be a stable solution, although it is a finite energy
solution. But this does not forbid us
from using the argument presented here to fix $\tan\beta$ because it does not
depend on the stability of the solution.
We hope future studies can clarify this issue.


\begin{flushleft}
{\sc Acknowledgements}: The author would like to thank
R. Arnowitt for helpful discussions. This work was supported in
part by NSF grant PHY89-07887 and World Laboratory.

\end{flushleft}

{\renewcommand{\Large}{\large}

}

%
%
%

\end{document}